\documentclass[prper,aps,twocolumn,groupedaddress,floats,showpacs,final,superscriptaddress]{revtex4-1}
\usepackage{leading}
\usepackage[latin3]{inputenc}
\usepackage[makeroom]{cancel}
\usepackage{graphicx}
\usepackage{amsmath}
\usepackage{amsfonts}
\usepackage{amssymb}

\usepackage{color}
\usepackage[left=2cm,right=2cm,top=2cm,bottom=2cm]{geometry}
\usepackage{graphicx} 
\usepackage{dcolumn}
\usepackage{bm}
\usepackage{simplewick}
\usepackage{array}
\usepackage{appendix}
\usepackage{ulem}

\RequirePackage[
   hyperindex,colorlinks,bookmarksnumbered,
   plainpages=true,pdfstartview=FitH]{hyperref}
\hypersetup{linkcolor=blue,urlcolor=blue,citecolor=blue}
\usepackage{hyperref}
\usepackage{mathrsfs}
\definecolor{purple}{rgb}{0.5,0,0.6}

\usepackage{ulem}
\renewcommand{\emph}[1]{\textit{#1}}
\definecolor{darkblue}{rgb}{0,0,0.5}
\definecolor{darkgreen}{rgb}{0,0.5,0}
\definecolor{darkred}{rgb}{.7,0,0}
\definecolor{purple}{rgb}{0.5,0,0.6}
\definecolor{orange}{rgb}{1,0.5,0}
\definecolor{grey}{rgb}{.6,.6,.6}
\definecolor{lightpink}{rgb}{1,0.7,0.75}
\definecolor{pink}{rgb}{1,0.4,0.58}
\definecolor{deeppink}{rgb}{1,0.08,0.58}

\newcommand{\DK}[1]{{\color{black}{#1}}} 
\newcommand{\DKK}[1]{{\color{black}{#1}}} 
\newcommand{\dpp}{\delta_{\rm P}}
\newcommand{\cp}{\cos2\delta_{\rm P}}
\newcommand{\cs}{\sin2\delta_{\rm P}}
\newcommand{\rss}{\mathbb{S}}

\renewcommand{\emph}[1]{\textit{#1}}

\newcommand{\Tk}{T_{\rm K}}
\newcommand{\mt}{\mathcal{T}}

\begin{document}

\title{Quantum thermoelectric and heat transport in the overscreened Kondo regime:\\ 
Exact conformal field theory results}

\author{D. B. Karki}
\affiliation{Division of Quantum State of Matter, Beijing Academy of Quantum Information Sciences, Beijing 100193, China}
\author{Mikhail N. Kiselev}
\affiliation{The  Abdus  Salam  International  Centre  for  Theoretical  Physics  (ICTP),
Strada  Costiera 11, I-34151  Trieste,  Italy}

\begin{abstract}
We develop a conformal-field theory approach for investigation of the quantum charge-, heat- and thermoelectric- transport through a quantum impurity fine tuned to a non-Fermi liquid regime. The non-Fermi-liquid operational mode is associated with the overscreened spin Kondo effect and controlled by the number of orbital channels. The universal low-temperature scaling and critical exponents for Seebeck and Peltier coefficients are investigated for the multichannel geometry. 
We discuss the universality of Lorenz ratio and power factor beyond the Fermi Liquid paradigm. Different methods of verifying our findings based on the recent experiments are proposed.
\end{abstract}

\date{\today}
\maketitle

\textit{Introduction.}$-$ The celebrated discovery of Kondo effect~\cite{Kondo} has led the development of different powerful techniques to tackled the weak and the strong coupling regimes of the Kondo model. Example includes perturbation theory~\cite{Kondo}, renormalization group~\cite{Anderson, *haman1, *Anderson_Yuval_Hamann, *Wilson}, Bethe ansatz~\cite{wigmann_JETP(38)_1983, Andrei_RevModPhys_1983} and conformal field theory~\cite{AFFLECK_NPB_1990,affadd, Al_prl, affadd1,Affleck_Lud_PRB(48)_1993, sengupta}. The complete understanding of the Kondo effect with spin-$1/2$ impurity fully screened by a single channel of conduction electrons, however, came after the seminal work of Nozieres~\cite{Nozieres} on the local Fermi-Liquid (FL) approach to the Kondo problem.

Shortly after achieving a breakthrough in understanding the single channel Kondo effect (1CK), Nozieres and Blandin (NB) put forth the physical picture of multi- $\mathcal{K}$-channel Kondo ($\mathcal{K}$CK) paradigm. The 1CK has then been extended for arbitrary spin $\mathcal{S}$ and arbitrary number of conduction channels $\mathcal{K}$~\cite{Nozieres_Blandin_JPhys_1980}. These generalized versions of Kondo model belong to fully screened for $\mathcal{K}=2\mathcal{S}$, an underscreened $\mathcal{K}<2\mathcal{S}$ and the overscreened $\mathcal{K}>2\mathcal{S}$ cases~\cite{Nozieres, Nozieres_Blandin_JPhys_1980,wigmann_JETP(38)_1983, *Andrei_RevModPhys_1983, *Sacramento_CM(48)_1991, *coleman_andrei, *Cox_Adv_Phys(47)_1998,  B_aff}. While the fully screened and the underscreened Kondo effects can be described within the FL paradigm, the overscreened case possess strikingly different behavior with a non-FL (NFL) properties~\cite{Nozieres}. Unlike the Dirac-electron-like quasiparticles as the low lying excitations in FL, those of 2CK are rather Majorana fermions and $\mathcal{K}$CK are believe to be described in terms of $Z_{\mathcal{K}}$ parafermionic excitations~\cite{Emery_PRB(46)_1992,*Maldacena_1997, *kane1, *ian1, *Ian_sela}. In addition, the  $\mathcal{K}$CK has recently been extended to the purely topological setup~\cite{Beri_Cooper_2012, *Altland_Egger_2013a, *Beri_2013b, *Galpin_Beri_2014a, *Tsvelik_2014e, *Beri_2017a, *yukka} suggesting the transport measurements in $\mathcal{K}$CK as an smoking gun for detecting Majorana zero modes.

The transport measurement in conventional 1CK has long history since its first observation~\cite{Goldhaber_nat(391)_1998} to the very recent experiment on the manipulation of the Kondo cloud~\cite{new}. Nevertheless, the hope for manipulating the exotic NFL quasiparticles has came just after the recent experimental success~\cite{Potok_NAT(446)_2007, Pierre_NAT(526)_2015, iff} on the observation of 2CK and 3CK physics. In addition, the richness of $\mathcal{K}$CK lies also on the fact that their spin version~\cite{Nozieres, Nozieres_Blandin_JPhys_1980, yuval_david,delft2, Potok_NAT(446)_2007, OGG, delft1, mitchell_sela} and charge version~\cite{mtv, Pierre_NAT(526)_2015, iff} behave differently. 

Apart from 1CK, the general understanding of quantum transport through quantum impurity in the presence of $\mathcal{K}$ orbital channels is limited to just the electronic conductance of prototypical NFL states. It is well know that thermoelectric response measurement at nano scale provides completely new sorts of information~\cite{Zlatic} tht can not be achieved by conductance measurement. On one hand, recent progress in the understanding of
thermoelectric phenomena in interacting nano systems has stimulated new experiments~\cite{kiselev, if3, *paulo, *heiner, new}. On the other hand, despite being widely studied for long time no single experiment has been yet reported for the thermoelectrics of $\mathcal{K}$CK effects. We believe that the recent experiments~\cite{Potok_NAT(446)_2007, Pierre_NAT(526)_2015, iff} on 2CK and 3CK would soon be extended for the thermoelectric measurement.

\begin{figure}[b]
\includegraphics[width=65mm,angle=0]{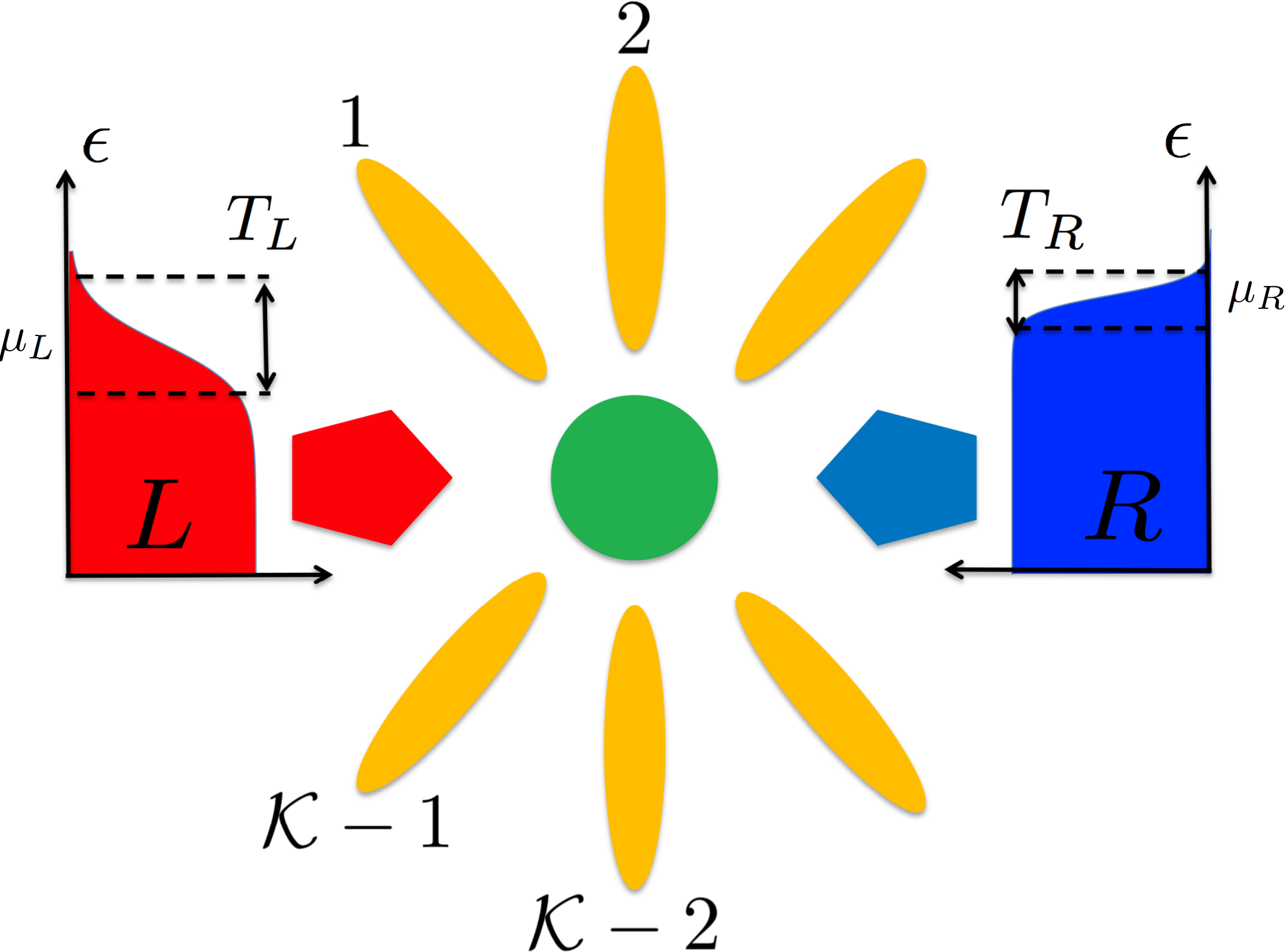}
\caption{{\color{black} Schematic setup for measurements of quantum transport in the overscreened $\mathcal{K}$-channel Kondo regime \cite{OGG}. Small quantum dot (green) is connected to $\mathcal{K}-1$ large metallic droplets (yellow) controlled by independent gates. 
The setup is connected to two metallic leads:
``hot" (red pentagon) characterized by temperature $T_L$ and chemical potential $\mu_L$ and 
``cold" (blue pentagon) with $T_R$ and $\mu_R$ correspondingly. For the thermoelectric measurements the voltage drop 
$\Delta$$V_{\rm th}$$=$$\mu_L$$-$$\mu_R$ is adjusted to compensate the electric current due to the temperature drop $\Delta$$T$$=$$T_L$$-T_R$.}}
\label{f1} 
\end{figure}

Despite the strong demand from the experimental community, 
full theory for quantum heat and thermoelectric transport in $\mathcal{K}$CK is still under construction ~\cite{moca, rok, mk, mkk}. In particular, the NFL thermoelectric properties of the {\it overscreened spin} Kondo problem is one of the most appealing of unsolved questions of the quantum transport through quantum impurity \cite{Potok_NAT(446)_2007}. In the light of recent experiments~\cite{Potok_NAT(446)_2007, Pierre_NAT(526)_2015, iff}, the case of paramount importance would be the spin 2CK and 3CK. 

In this Rapid communication we will focus on thermoelectric and heat transport through the quantum devices investigated in \cite{Potok_NAT(446)_2007}.
The setup \cite{Potok_NAT(446)_2007} fabricated for studying electric conductance 
in the regime of two-channel spin Kondo problem consists of two electron droplets (Quantum Dots, QD): one large and one small. Stabilization of the two-channel strong coupling fixed point is achieved by adjusting the electrochemical potential in each droplet. This setup can straightforwardly be used for thermoelectric measurements by creating a temperature drop across the small QD. Generalization for multi-channel spin Kondo problem can be done by inclusion of several ({\color{black}$\mathcal{K}-1$}) large QDs with adjustable charging energy (see Fig.\ref{f1}). In the following, thus, we restrict ourselves by the investigation of fully-fledged theory of thermoelectrics at the strong coupling regime of overscreened spin $\mathcal{K}$CK effects.

\textit{Model.}$-$ The Hamiltonian describing the exchange coupling between the $\mathcal{K}$ degenerate channels of spin-1/2 conduction electrons and the impurity with an effective spin $\mathcal{S}$ reads~\cite{Nozieres_Blandin_JPhys_1980}
\begin{equation}\label{aama1}
\mathscr{H}{=}{\sum_{k}}\varepsilon_k \left(\psi_k^{\alpha, i}\right)^{\dagger}\psi_{k\alpha, i}{+}J{\bf \mathcal{S}}\cdot\sum_{kk'}\left(\psi_k^{\alpha, i}\right)^{\dagger}
\frac{\sigma_{\alpha}^{\beta}}{2}\psi_{k'\beta,i},
\end{equation}
where $\alpha, \beta=\uparrow,\downarrow$ and $i{=}1, 2,\cdots\mathcal{K}$ stand for spin and channel indices respectively and $\sigma_{\alpha}^{\beta}$ are the Pauli matrices acting in spin sector. The operator $\psi_{k\alpha,i}$ annihilates an electron in the $k\alpha$ state of the conduction channel $i$. The strength of exchange interaction is accounted for by the parameter $J$. In addition, the summation over repeated raised and lowered indices is implied in Eq.~\eqref{aama1}

The scattering matrix provides tentative idea on the low energy properties of the model~\eqref{aama1}. Using boundary conformal-field-theory (BCFT) approach, Affleck and Ludwig (AL) computed a general expression of the scattering matrix amplitude $\rss$ for an incoming electron to scatter into one electron at the Fermi surface and at zero temperature
\begin{equation}\label{aama2}
\rss=\cos \left[\frac{\pi\left(2\mathcal{S}+1\right)}{\mathcal{K}+2}\right]\Big/
\cos \left[\frac{\pi}{\mathcal{K}+2}\right],
\end{equation}
In fully screened case the resulting unitarity condition $|\rss|{=}1$ signifies the trivial $\pi/2$ phase shift between incoming and outgoing electron states. In contrast, $|\rss|{<}1$ for $\mathcal{K}{\geq}2$ shows the strengthening multi-particle scattering events in NFL state even at zero temperature ($T{=}0$). From Eq.~\eqref{aama2} it is seen that the general setups satisfying the condition $\mathcal{K}=4\mathcal{S}$ have vanishing $\rss$. This results in the complete absence of single particle scattering events at zero temperature.

The multi-channel Kondo Hamiltonian Eq.~\eqref{aama1} possess an exact $SU(\mathcal{K})$ channel symmetry on the top of an exact $SU(2)$ spin symmetry and $U(1)$ charge symmetry. Consequently, the low energy observables of the model~\eqref{aama1} are particle-hole (PH) symmetric. While NFL fixed point is unstable for certain perturbations, such as magnetic field and channel anisotropy, the PH symmetry can trivially be broken by potential scattering without affecting the Kondo physics~\cite{Affleck_Lud_PRB(48)_1993}. Since the potential scattering acts merely in the charge sector, the leading irrelevant operator describing the Kondo physics which lives entirely in the spin sector has no real effect of potential scattering, for weak Kondo coupling. The effects of potential scattering can be accounted for by considering an additional term $\delta \mathscr{H}_{\rm P}$ in the one dimensional left moving (incoming) theory~\cite{Affleck_Lud_PRB(48)_1993}
\begin{equation}\label{aama3}
\delta \mathscr{H}_{\rm P}=\frac{v\;\delta_{\rm P}}{\pi}\mathcal{J}_{\rm L}(0),
\end{equation}
where $\mathcal{J}_{\rm L}(0)$ is the charge current \DK{(see Ref.~\cite{AFFLECK_NPB_1990} for details)}, $v$ is the velocity of left moving fermion field and $\delta_{\rm P}$ is a constant phase shift produced by the potential scattering.

\textit{Self energy and T-matrix.}$-$ In AL BCFT approach, the energy and temperature-dependent retarded self energy $\Sigma^{\rm R}(\varepsilon, T)$ coming from perturbation in leading irrelevant operator (LIO) gets simply multiplied by the factor $e^{2i\delta_{\rm P}}$ due to the inclusion of $\delta \mathscr{H}_{\rm P}$. \DK{At the strong coupling regime $T\ll\Tk$, with $\Tk$ being the corresponding Kondo temperature, the self energy reads}~\cite{Affleck_Lud_PRB(48)_1993}
\begin{align}\label{aama4}
{\Sigma^{\rm R}(\varepsilon, T)}={{-}{\frac{i}{2\pi\nu}}}{\Big[}1{-}\rss\; e^{2i\delta_{\rm P}}{-}e^{2i\delta_{\rm P}}\mathscr{M}\lambda \left(\pi T\right)^{\Delta}\mathcal{I}{\Big]}.
\end{align}
In Eq.~\eqref{aama4}, $\nu$ is the density of states per spin per channel, $\lambda$ is leading irrelevant coupling constant and $\Delta$ is the dimension of the boundary operator which defines the dimension of corresponding LIO $\mathscr{O}$ at NFL fixed point such that $\mathscr{O}{=}1+\Delta$ with
\begin{eqnarray}
\Delta\equiv \frac{2}{\mathcal{K}+2}.
\end{eqnarray}
In addition, the symbol $\mathscr{M}$ in Eq.~\eqref{aama4} is the function of $\Delta$ and the effective spin of impurity $\mathcal{S}$ \footnote{\DKK{The consideration of an arbitrary spin $\mathcal{S}$ affects only the amplitude of the scattering matrix $\mathbb{S}$ and the numerical factor $\mathscr{M}$ as long as the condition of overscreening $\mathcal{K}>2\mathcal{S}$ is satisfied~\cite{Affleck_Lud_PRB(48)_1993}.}}. In this work we restrict ourself by considering the case of $\mathcal{S}=1/2$ for which $\mathscr{M}$ reads~\cite{Affleck_Lud_PRB(48)_1993,Gogolin_book}
\begin{eqnarray}\label{MK}
\mathscr{M}{=}
3\cdot 2^{\Delta}\sin\left(\pi \Delta\right)
\sqrt{\!\frac{{\sin \left(\frac{\pi \Delta}{2}\right)}{ \tan \left(\frac{\pi \Delta }{2}\right)}{ \Gamma \left(1-\Delta\right)^2}}
{\Gamma \left(1-3\Delta/2\right)\Gamma \left(1-\Delta/2\right)}}.\;\;\;\;
\end{eqnarray}
Here $\Gamma$ is Euler's gamma function and $\mathcal{I}$ represents an integral
\begin{align}\label{aama5}
\mathcal{I}\equiv& {\int^{1}_{0} du}\Big[u^{-\frac{i\varepsilon}{2\pi T}} u^{{-\frac{1}{2}}}(1{-}u)^{\Delta} F(u, \Delta){-}\mathcal{P}(u, \Delta)\Big],
\end{align}
where we use short-hand notations for the hypergeometric function $_2F_1$$($$1$$+$$\Delta,$$ $$1$$+$$\Delta$$;$$1$$;$$u)$$\equiv$$F$$(u,$$\Delta$$)$
\begin{align}\label{aama6a}
F(u, \Delta)&=\frac{1}{2\pi}\int^{2\pi}_{0}\frac{d\theta}{(u+1-2\sqrt{u}\cos\theta)^{1+\Delta}},
\end{align}
and the function $\mathcal{P}(u,\Delta)$ is defined as 
\begin{align}\label{aama6b}
\mathcal{P}(u,\Delta)&\equiv\; \frac{\Gamma[1+2\Delta]}{\Gamma^2[1+\Delta]}\frac{u^{\Delta-1}}{(1-u)^{\Delta+1}}.
\end{align}
For the case of single quantum impurity under consideration, the self energy and the scattering T-matrix $\mt(\varepsilon, T)$ are connected via $\Sigma^{\rm R}(\varepsilon, T)=\mt(\varepsilon, T)$. The quantity defined by the imaginary part of $-\pi\nu\mt(\varepsilon, T)$ then defines the spectral function, the central object of transport calculation.

\textit{Thermoelectric transport coefficients.}$-$ To proceed with the \DK{linear response} thermoelectric transport calculation in overscreened $\mathcal{K}$CK regime, we define the transport integrals~\cite{casti} 
\begin{equation}\label{aama8}
\mathscr{L}_{\rm n}\equiv\frac{1}{4T}\int^{\infty}_{-\infty}\;\frac{d\varepsilon}{\pi}\;\frac{\varepsilon^{\rm n}\cdot{\rm Im}\left[-\pi\nu\mathcal{T}(\varepsilon, T)\right]}{\cosh^2(\varepsilon/2T)},
\end{equation}
characterizing thermoelectric transport properties \footnote{The linear response thermoelectric measurements are fully accessible via the transmission coefficient evaluated at equilibrium~\cite{casti}}. Note that in Eq.~\eqref{aama8} and throughout this paper we are using the system of atomic unit $e=\hbar=k_{\rm B}=1$ unless otherwise stated explicitly. The coefficient $\mathscr{L}_0$ provides an access to the electrical conductance $G$, the thermoelectric coefficient $G_{\rm T}$ is defined in terms of $\mathscr{L}_1$ such that $G_{\rm T}=\mathscr{L}_1/T$. The other measure of thermoelectric transport such as the thermopower $\mathcal{S}_{\rm th}$ and the electronic thermal conductivity $\mathscr{K}$ are also defined in terms of $\mathscr{L}_{\rm n}$~\cite{casti}
\begin{equation}\label{aama9}
\mathcal{S}_{\rm th}(T)=\frac{1}{T}\frac{\mathscr{L}_1}{\mathscr{L}_0},\;\;\mathscr{K}(T)=\frac{1}{ T}\left[\mathscr{L}_{\rm 2}-\frac{\mathscr{L}_{\rm 1}^2}{\mathscr{L}_{\rm 0}}\right].
\end{equation}
In addition, the Wiedemann-Franz (WF) law connects the electronic thermal conductivity $\mathscr{K}$ to the electrical conductance $G$ in low temperature regime of a Fermi liquid by an universal constant, the Lorenz number $L_0$, defined as $L_0\equiv \mathscr{K}_{\rm FL}/G_{\rm FL}T=\pi^2/3$. Although WF law naively implies that the transport mechanisms responsible for heat and charge currents are fundamentally the same, surprisingly it works quantitatively well even for some interacting nano systems~\cite{costi1}. The possible deviation from WF in NFL is accounted for by studying the Lorenz ratio $\mathscr{R}\equiv L(T)/L_0$ where $L(T)=\mathscr{K}/GT$ computed in the NFL regime.

To get the transport integrals, we first plug in the spectral function obtained from Eq.~\eqref{aama4} into the expression of transport integrals Eq.~\eqref{aama8}. Afterward by changing the order of integration between the variables $\varepsilon$ and $u$, we integrated out exactly the variable $\varepsilon$. The transport integrals are then expressed in terms of the integral over $u$
\begin{align}\label{aama10}
\mathscr{L}_{\rm 0}&=\frac{1}{2\pi}\Big(1-\rss\cp+\mathscr{M}\mathscr{C}_0\cp\;\lambda(\pi T)^{\Delta}\Big),\nonumber\\
\mathscr{L}_1&=\frac{1}{2\pi}\Big(\mathscr{M}\;\mathscr{C}_1\;\cs\;\lambda(\pi T)^{\Delta+1}\Big),\\
\mathscr{L}_2&=\frac{1}{6\pi}(\pi T)^2\Big(1{-}\rss\cp {+}3\mathscr{M}\mathscr{C}_2\cp\;\lambda (\pi T)^{\Delta}\Big).
\nonumber
\end{align}
The coefficients $\mathscr{C}_n$ in the expression of $\mathscr{L}_n$ are given by
\begin{align}\label{aama11}
\mathscr{C}_0 (\Delta)&=\int^{1}_{0}du\Big[\frac{F(u, \Delta)\;\log(u)}{(1-u)^{1-\Delta}}+\mathcal{P}(u, \Delta)\Big].\\
\mathscr{C}_1(\Delta)&= \int^{1}_{0}du\; F(u,\Delta)\;\frac{2 (u-1)-(u+1) \log (u)}{(1-u)^{2-\Delta}},\nonumber\\
\mathscr{C}_2(\Delta)&={\int^{1}_{0}}\!du \Big[\!{{-}F(u, \Delta)}\frac{4(1{-}u^2){+}\{u (u{+}6){+}1\} \log (u)}{(1-u)^{3-\Delta}}\nonumber\\
&\;\;\;\;\;\;\;\;\;\;\;\;\;\;\;\;\;\;\;\;\;\;\;\;\;\;\;\;\;\;\;+\frac{1}{3}\;\mathcal{P}(u, \Delta)\Big].\nonumber
\end{align}
\DK{Using a combination of both analytic and numerical tools~\footnote{\DKK{Although we have not yet been able to prove this equations analytically, the extensive numerical calculations with high precision goal demonstrate that the equations are exact for $\Delta$ defined in the interval 
$\Delta\in (0,1/2]$. We have numerically checked our predictions explicitly for $\mathcal{K}=2-100$ and the difference between the numerical calculation and the corresponding quantities expressed in Eq.~\eqref{kd2} are found to be within the numerical error
of integral computations associated with Mathematica package. These details will be published somewhere else.}}, we obtain simple form for $\mathscr{C}_{0, 1, 2}$, in terms of rational functions of $\Delta$}
\begin{eqnarray}\label{kd2}
&&\mathscr{C}_0(\Delta)=\frac{1}{\Delta(1+\Delta)},\;\;\;\mathscr{C}_1(\Delta)=\frac{1}{(1-\Delta)(2+\Delta)},\nonumber\\
&&\mathscr{C}_2(\Delta)={\color{black}\frac{1}{3}\left[\frac{1}{\Delta(1+\Delta)}
+\frac{4}{(2-\Delta)(3+\Delta)}\right]}.
\end{eqnarray}
\DKK{This prediction can easily be verified by the method of large $\mathcal{K}$ expansion}. 
Similar prediction for the dependence of transport coefficients on the number of conduction channels in FL regime has been appeared in Ref.~\cite{HWDK_PRB_(89)_2014}. 

\textit{Large $\mathcal{K}$-limit}$-$ As noticed by NB that the large $\mathcal{K}$-limit of overscreened $\mathcal{K}$CK results in a rather distinct fixed point where LIO $\mathscr{O}$ possess a non-trivial scaling dimension of $1+2/\mathcal{K}$~\cite{Nozieres_Blandin_JPhys_1980}. Interestingly, this non-trivial critical behavior can be consistently explained in terms of the perturbative expansion in $1/\mathcal{K}$~\cite{Affleck_Lud_PRB(48)_1993}. The $\mathcal{K}$-dependence on the Onsager transport coefficients expressed in Eq.~\eqref{aama10} comes from the functions $\rss$ and $\mathscr{M}\mathscr{C}_n$. Therefore to get further insights on the behavior of $\mathscr{L}_n$ for $\mathcal{K}\to \infty$, we perform the $1/\mathcal{K}$ expansion of $\rss$ and $\mathscr{M}\mathscr{C}_n$. These expressions in leading and sub-leading order in $1/\mathcal{K}$ are given by
\begin{align}\label{kap1}
& \rss  =1-\frac{3\pi^2}{2}\;\frac{1}{\mathcal{K}^2}+\mathcal{O}\left(\frac{1}{\mathcal{K}^3}\right),\;\mathscr{M}\mathscr{C}_1=\frac{3\pi^2}{\mathcal{K}^2}+\mathcal{O}\left(\frac{1}{\mathcal{K}^3}\right),\nonumber\\
&\mathscr{M}\mathscr{C}_0  =\frac{3\pi^2}{\mathcal{K}}+\frac{6\pi^2\;\left[\log(2)-2\right]}{\mathcal{K}^2}+\mathcal{O}\left(\frac{1}{\mathcal{K}^3}\right),\nonumber\\
& \mathscr{M}\mathscr{C}_2=\frac{\pi^2}{\mathcal{K}}+\frac{2\pi^2\;\left[\log(2)-4/3\right]}{\mathcal{K}^2}+\mathcal{O}\left(\frac{1}{\mathcal{K}^3}\right).
\end{align}
{\color{black} These asymptotes} adequately explain the charge and energy transfer processes in overscreened Kondo effects with large number of conduction channels.

\textit{Main results.}$-$ From the expression of $\mathscr{L}_0$ we cast the expression of conductance in $\mathcal{K}$CK regime 
\begin{equation}\label{aama12}
G(T)=G_0\left[1{-}\frac{\mathscr{M}\mathscr{C}_0\cp}{1-\rss\cp}\!\left(\frac{\pi T}{\Tk}\right)^{\Delta}\right],
\end{equation}
where the zero temperature conductance is defined as 
$G_0$$\equiv $$\left(\right.$$1{-}$$\rss$$\cp$$\left.\right)$$/$$2$$\pi$. In addition we define the Kondo temperature $\Tk$ via the relation $\lambda\equiv -\Tk^{-\Delta}$~\footnote{In general the leading irrelevant coupling constant $\lambda$ can take either sign (see detailed discussion in~\cite{Affleck_Lud_PRB(48)_1993}). Thus, contrast to FL regime where the temperature behavior of transport coefficients is determined by the $\lambda^2$ corrections, the transport coefficients in the NFL regime is characterized by either increasing or decreasing behavior. However, the experiments~\cite{Potok_NAT(446)_2007} performed in 2CK regime demonstrated that the electric conductance decreases with increasing the temperature and thus $\lambda$ is negative (see Ref.~\cite{OGG} for details)}. The last equation expresses the ${\sim} T^{\Delta}$ scaling of the conductance for $\mathcal{K}$CK effects with two parameters, the potential scattering $\dpp$ and the Kondo temperature $\Tk$~\footnote{In the large $\mathcal{K}$ limit of the overscreened Kondo regime the coefficient in front of the temperature dependent term of electric conductance is proportional to $\mathcal{K}$ similarly to FL fully screened regime analyzed in~\cite{HWDK_PRB_(89)_2014}. The leading term in the $1/\mathcal{K}$ expansion for both Seebeck and Peltier coefficients is $\mathcal{K}$-independent}.

\textit{Seebeck effect.}$-$ The thermoelectric power (Seebeck coefficient) is determined by $\mathscr{L}_0$ and $\mathscr{L}_1$ (see Eq.~\ref{aama9}). Plugging in the transport integrals into Eq.~\eqref{aama9} \DK{we obtain a compact expression of thermopower at $T\ll\Tk$}
\begin{equation}\label{aama13}
\mathcal{S}_{\rm th}(T)=-\frac{\pi\mathscr{M}\;\mathscr{C}_1\;\cs\;\left(\frac{\pi T}{\Tk}\right)^{\Delta}}{1-\rss\cp-\mathscr{M}\mathscr{C}_0\cp\left(\frac{\pi T}{\Tk}\right)^{\Delta}}.
\end{equation}
The Eq.~\eqref{aama13} shows that the low temperature behavior of thermopower in overscreened Kondo effects also scales as ${\sim}T^{\Delta}$ as an smoking gun signature of NFL correlation \footnote{\color{black} The scaling of thermopower $\mathcal{S}_{\rm th}\propto T^\Delta$ is different from the corresponding scaling of the entropy $\left(S_{\rm ent}-S_0\right) \propto T^{2\Delta}$. Note that for $K=2$, $\Delta=1/2$ the scaling is
$\left(S_{\rm ent}-S_0\right)\propto T\ln T$. Besides, contrast to Seebeck coefficient, the entropy at $T=0$ limit acquires residual value 
$S_0=\ln\left[2\cos(\pi\Delta/2)\right]$ related to degeneracy of the ground state~\cite{affadd, Al_prl, affadd1}}. It is also noted that coefficient $\mathscr{C}_1$ in Eq.~\eqref{kd2} controlling the leading order behavior of thermopower vanishes quickly with increasing $\mathcal{K}$. This particular behavior indicates that there is indeed the bottleneck signature for the energy transport unlike the charge transport. This might be because at low energy regime only few channels coupled with impurity degree of freedom and rest of them get effectively decoupled narrowing the window for energy transport. By its very relation with the entropy, thermopower expression Eq.~\eqref{aama13} also opens a viable door for exploring the entropy production in NFL regime~\cite{erik, meir2019, sela_folk}. Of the particular experimental interest, we explicitly write the thermopower of 2CK which casts into the compact form~\footnote{Negative sign of $\lambda$ determines the sign of thermopower at fixed particle-hole asymmetry defined by the gate voltage through $\delta_{\rm P}$}
\begin{equation}\label{aama14}
 \mathcal{S}^{\rm 2CK}_{\rm th}=-\frac{12\pi}{5}\sin 2\dpp\sqrt{\frac{\pi T}{\Tk}}+ 
\mathcal{O}\left(\sin 4\dpp\cdot\frac{\pi T}{\Tk}\right).
\end{equation}
This compelling form Eq.~\eqref{aama13} mainly results from the fact that the S-matrix vanishes at 2CK overscreened regime ($\rss=0$ in the Eq.~\eqref{aama2} for $\mathcal{K}=2$). 

\textit{Peltier effect.}$-$  Peltier coefficient $\Pi$  characterizes the generation of heat current $I_{\rm h}$ due to the charge current $I_{\rm c}$ driven in a circuit under isothermal condition by an applied voltage bias~\cite{ceo1}. This coefficient provides a valuable information on the characterization of how good a material is for thermoelectric solid-state refrigeration or power generation. The expression of thermopower Eq.~\eqref{aama13} directly connects the Peltier coefficient via the Kelvin relation $\Pi=T\mathcal{S}_{\rm th}$. Therefore the leading order temperature dependence of Peltier coefficient $\Pi\sim T^{1+\Delta}$ has markedly different behavior compared to the corresponding FL case $\Pi_{\rm FL}\sim T^2$.

\textit{Power factor.}$-$
Another quantity of paramount experimental importance that can be directly read out from $G$ and $\mathcal{S}_{\rm th}$ is the power factor $Q\equiv \mathcal{S}^2_{\rm th} G$~\cite{whitney1}. In conventional FL situation since $\mathcal{S}^{\rm FL}_{\rm th}\sim T$ and $G^{\rm FL}-G^{\rm FL}_0\sim T^2$, the power factor follows the low-T scaling law $Q_{\rm FL}\sim T^2$~\cite{dee1, *last1, *last2} which, thus, limited to very small value. From our previous discussions on the scaling behavior of $G$ and $\mathcal{S}_{\rm th}$, the $Q\propto T^{2\Delta}$ scaling of the power factor can be observed.  As a result, the power factor in NFL regime can be significantly higher then that in FL regime of Kondo effects.

\textit{Wiedemann-Franz law.}$-$
From the direct observation of the temperature scaling from of transport integrals in Eq~\eqref{aama10}, it is evident that the WF law satisfies even in overscreened Kondo regime at the low temperature limit $T$$\to $$0$. As an example, the straightforward calculation of Lorenz ratio results in the leading temperature dependence in 2CK regime as $\mathscr{R}_{\rm 2CK}$$\simeq $$1{-}1.29$$\cp $$\sqrt{\pi T/\Tk}$ which satisfies WF law for $T$$\to $$0$. 

\textit{Discussion and open questions.}$-$ The key predictions of this Rapid communication can be directly verified using existing experimental setup~\cite{Potok_NAT(446)_2007}. To perform the thermopower measurements it is necessary to heat one of the leads (e.g. with Joule heat) to create controllable temperature drop $\Delta T$ across the QD. 
Measurement of  the Seebeck coefficient $\mathcal{S}_{\rm th}(T)=-\Delta V_{\rm th}/\Delta T$ is achieved by applying a {\it thermo-voltage} $\Delta V_{\rm th}$ nullifying  the net current through the nano-device. Connection between the Seebeck and Peltier coefficients $\Pi=T\mathcal{S}_{\rm th}$ provides an experimental tool to control the heat current through the device operating in the NFL regime. It, in turn, gives an access to experimental measure of the Lorenz ratio for verification of the Wiedemann-Franz law. 

Yet another theoretical challenge is to apply the conformal-field theory approach for analyzing a shot noise in the NFL regime by computing a current-current correlation function. The information about quantum noise gives a direct access to the charge fractionalization and can be experimentally investigated using existing techniques~\cite{Pierre_NAT(526)_2015}. In addition, theoretical analysis of {\it full-counting statistics} is necessary to shed the light on temperature scaling of arbitrary order of the charge current cumulants to determine the {\it moment-generating function} for  the  probability  distribution  function of transferred  charges  within  a  given  time  interval~\cite{Gogolin_Komnik_PRB(2006), sela1}. We believe that the theoretical approach developed and reported in this Rapid communication is the necessary step to find an exact solution of these long-standing problems.

\textit{Conclusions.}$-$ In this Rapid communication we constructed a full fledged theory of the quantum thermoelectric and heat transport in the NFL regime associated with the {\it overscreened} multichannel behavior of spin Kondo impurity. By using exact conformal-field theory methods free of any additional assumptions (e.g.~Toulouse point etc. \cite{Gogolin_book}) we analyzed the low-temperature scaling of the Seebeck and Peltier coefficients and computed Lorenz ratio. The temperature scaling of thermoelectric and heat transport coefficients demonstrates pronounced deviation from conventional behavior characteristic for the FL regime. 
{\color{black} We suggest to verify theoretical predictions summarized in this Rapid communication using multi-Quantum-Dot setups \cite{Potok_NAT(446)_2007}.} The experimental verification of the Wiedemann-Franz law and, in particular, the finite-temperature corrections to it's zero-temperature limit is suggested to be used as an additional benchmark for the quantum transport measurements. It is predicted that the {\it power factor} in the NFL regime of the {\it overscreened} Kondo model can be significantly higher compared to the corresponding FL behavior of the reference {\it fully-screened} Kondo system opening an avenue for practical applications of the NFL quantum transport.

\textit{Acknowledgements.}$-$ The work of M.K. was supported in part by the National
Science Foundation under Grant No. NSF PHY-1748958 and conducted within
the framework of the Trieste Institute for Theoretical Quantum
Technologies (TQT).
%
\end{document}